\documentclass[12pt,twoside,a4paper]{amsart}
\usepackage[dvipsnames]{xcolor}                     
\usepackage{hyperref} 
\hypersetup{breaklinks, colorlinks,                 
    linkcolor = {Blue},                             
    citecolor = {Blue},                             
    urlcolor  = {Blue}                              %
}                                                   %
\newcommand{\urlp}[1]{\hypersetup{urlcolor = {BrickRed}}\url{#1}\hypersetup{urlcolor = {Blue}}} %
\usepackage[OT2, T1]{fontenc}\usepackage[russian, english]{babel}
\usepackage[T1]{fontenc}
\usepackage[latin9]{inputenc}
\usepackage[margin=20mm,top=30mm,bottom=30mm]{geometry}
\usepackage{amsmath,amssymb,amsthm,amscd,stmaryrd,graphicx,bm,latexsym}
\usepackage{mathrsfs}\usepackage{mathtools}\usepackage{lmodern}\usepackage{natbib}\usepackage{enumitem}\usepackage{soul}
\newlist{parts}{enumerate}{5}       
\setlist[parts]{wide,label=\textbf{\upshape(\alph*)},ref=(\alph*),align=left,labelindent=0pt,labelsep=.5em,itemsep=.3ex,topsep=0ex,%
}
\usepackage[
final,
scrtime
]{prelim2e}

\renewcommand{\PrelimText}{\footnotesize[\,Version: 
\texttt{\jobname.tex}\hfill \today\ at \thistime\,]}
  \theoremstyle{plain}
    \newtheorem{Thm}{Theorem}[section]  
      \newtheorem{Lem}[Thm]{Lemma}
   
    \theoremstyle{definition} 

      \newtheorem{Remarks}[Thm]{Remarks}
\usepackage{amsthm}
\usepackage{mdframed} 

\usepackage{scalerel}
\newcommand{\pr}[1]{\scalerel*{\textbf{(}}{\strut}#1\scalerel*{\textbf{)}}{\strut}}

\renewcommand{\iff}{\Leftrightarrow}

\renewcommand{\cite}{\citet}
\newcommand*{\nameadjunct}{\relax}
\makeatletter
\renewcommand*{\NAT@nmfmt}[1]{\NAT@up #1\nameadjunct}
\makeatother

\newcommand*{\citeposs}[2][]{%
    \begingroup
    \renewcommand*{\nameadjunct}{'s}%
    \citet[#1]{#2}%
    \endgroup
}

\newcounter{aequiv}

 {\end{list}} 
\let\oldsqrt\sqrt
\def\sqrt{\mathpalette\DHLhksqrt}
\def\DHLhksqrt#1#2{%
\setbox0=\hbox{$#1\oldsqrt{#2\,}$}\dimen0=\ht0
\advance\dimen0-0.2\ht0
\setbox2=\hbox{\vrule height\ht0 depth -\dimen0}%
{\box0\lower0.4pt\box2}}

\newcommand{\N}{{\mathbb N}}\newcommand{\R}{{\mathbb R}}%
\newcommand{\Z}{{\mathbb Z}}
\newcommand{\cP}{{\mathcal P}}%
\newcommand{\cX}{{\mathcal X}}%
\renewcommand{\epsilon}{\varepsilon}\renewcommand{\phi}{\varphi} %
\renewcommand{\rho}{\varrho}\renewcommand{\theta}{\vartheta}     %
\DeclareMathOperator{\sgn}{sgn}  
\DeclareMathOperator*{\bigconv}{\mbox{\LARGE$\ast$}}
\DeclareFontEncoding{LS2}{}{\noaccents@}                                                 %
\DeclareFontSubstitution{LS2}{stix}{m}{n}                                                %
\DeclareSymbolFont{largesymbols_stix}{LS2}{stixex}{m}{n}                                 %
\DeclareMathDelimiter{\lpp}{\mathopen}{largesymbols_stix}{"DE}{largesymbols_stix}{"02}   %
\DeclareMathDelimiter{\rpp}{\mathclose}{largesymbols_stix}{"DF}{largesymbols_stix}{"03}  %
\renewcommand{\[}{\begin{eqnarray*}}\renewcommand{\]}{\end{eqnarray*}}
\newcommand{\la}{\begin{eqnarray}}\newcommand{\al}{\end{eqnarray}}
  
\renewcommand{\Prob}{\mbox{\rm Prob}}      
\newcommand{\dd}{\mathrm{d}}

\newcommand{\sdot}{\bullet}  

\begin{document}
\author[Mattner]{Lutz Mattner}
\title[Teachable normal approximations to binomial and related]%
{Teachable normal approximations to \\
  binomial and related \\
 probabilities  or confidence bounds 
 }
\begin{abstract}
For the usual normal approximations to binomial, hypergeometric, or Poisson interval probabilities,
we collect some simple but then reasonably sharp error bounds.
For the Clopper-Pearson~(1934) binomial confidence bounds,
we present, following Michael Short's~(2023) approach, bounds similar to,
but necessarily more complicated than, Lagrange's (1776) success rate
plus/minus normal quantile times estimated standard deviation.

The bounds, as presented here in four theorems,
should be teachable, to
people ranging from sufficiently advanced high school pupils
to university students in mathematics or statistics:  For understanding  most
of the proposed approximation results, it should suffice to know binomial laws,
their means and variances, and the standard normal distribution function,
but not necessarily the concept of a corresponding normal random variable.

Accompanying technical remarks, references, and proofs
are meant for assuring teachers or for stimulating further research.

Of the proposed approximations, some are essentially well-known at least to experts,
and some are based on teaching experience and research at Trier University.
\end{abstract}
\date{\today} 
\maketitle

\setcounter{section}{-1}
\section{Introduction}
The central limit theorem specialised to binomial random variables
$X_{n,p}$ with the indicated parameters says that the distribution
functions 
$\widetilde{F}_{n,p}$ of the standardised random variables
$\widetilde{X}_{n,p} \coloneqq \left(X_{n,p}-np\right)/\sqrt{np(1-p)}$
converge for $n\rightarrow \infty$ to the standard normal distribution function $\Phi$,
namely
\begin{equation}                    \label{Eq:binomial_CLT}
 \lim_{n\rightarrow\infty} \sup_{x\in\R} \left| \widetilde{F}_{n,p}(x)-\Phi(x)\right| \ =\ 0
 \quad\text{ for each }p\in\mathopen]0,1\mathclose[\,.
\end{equation}
This result is hardly interpretable in any practically meaningful way,
first since mere limit results never are, and second since the convergence is not
uniform in the parameter $p$. The latter fact is particularly troublesome when
trying to use~\eqref{Eq:binomial_CLT} for
constructing confidence bounds for $p$, or when trying to approximate given confidence
bounds like Clopper-Pearson's.

Hence, certainly for university students of mathematics or statistics, but
also for sufficiently advanced high school pupils, the above limit result should be replaced
or at least accompanied with error bounds or other explicit inequalities.

There are many possibilities of such inequalities,
some too imperfect or too ugly  for almost everybody,
some attractive to experts but
too complicated for most students, let alone high school pupils,
some as yet only conjectured but not proven,
and some still to be found in future mathematical research.

Here we aim
at presenting, in Theorems~\ref{Thm:Hypergeometric}, \ref{Thm:Binomial}, \ref{Thm:Symmetric_binomial}, and
\ref{Thm:CP-bounds_bounded},
some arguably simple,
elegant, useful,  proven, and hence teachable,
inequalities, mainly for binomial laws.
The accompanying remarks explain possible refinements,
optimalities,
or limits to conjectures of improvements.
Some of these results are essentially well-known at least to experts,
some are based at least partly on research at Trier University,
in particular on the dissertations of \citet{Schulz2016} and \citet{vanNerven2024},
and some are based on our teaching, in case of Theorem~\ref{Thm:CP-bounds_bounded}
following the approach of \citet{Short2023}.

We have tried to state the theorems  in a way to be as
widely teachable as possible, perhaps after some obvious
simplifications depending on the audience, for example:
In Theorem~\ref{Thm:Hypergeometric} one might omit the
subclause about~\eqref{Eq:lower_concentration_variance_bound}
entirely or  one might state just one of the two lower bounds there,
and one might replace the numerical constant $0.6879$
by the simpler but rougher~$0.7$\,.

By contrast, the remarks following the theorems present various
details and clarifications, of which perhaps in each case
the first few are still quite widely teachable,
whereas others might only be of interest for anybody trying
to improve the theorems,
or for the teacher of the theorems to make sure he knowns what he is talking about.

All proofs are collected in sections~\ref{sec:Proofs_1--3} and~\ref{sec:Proofs_4}.
They consist to a large part of references to the literature.

In the theorems, we have tried to avoid any not strictly necessary notation,
using essentially only $\Phi$ for the standard normal distribution function,
$\Phi^{-1}$ for its inverse (or quantile) function,
and ``$\mathrm{L.H.S.}$'' for ``left hand side''.
In the rest of the paper, we freely use further more less standard
mathematical notation and jargon,
such as ``iff'' for ``if and only if'',  and
the supremum norm notation $\left\|H\right\|_\infty \coloneqq \sup_{x\in\R}|H(x)|$.
If $P$ is a law on $\R$,
we write $\mu(P)$ for its mean (if existent)
and $\sigma(P)$ for its standard deviation,
just $F$ for its distribution function,
and if $0<\sigma(P)<\infty$ also $\widetilde{F}$ for the standardisation
of $F$, namely $\widetilde{F}(x)\coloneqq F\big(\sigma(P)x+\mu(P)\big)$ for $x\in\R$.
\pr{Parantheses as around this sentence indicate footnote-like
remarks one might just skip.}

\section{Normal approximations to hypergeometric interval probabilites,
and hence also to binomial and Poisson ones}   \label{sec:Hypergeometric}
After binomial laws,
every introduction to probability or statistics will sooner or later
lead to hypergeometric and Poisson laws. It is fortunate that,
for these three kinds of laws, a quite sharp normal approximation error bound
can be given in a simple and unified way. Very roughly speaking, such a law is approximately normal iff its standard deviation is large.
A bit more precisely, with $\sigma$ denoting the standard deviation of the law in question,
the normal approximation error for (general) interval probabilities
is of the order $\frac{1}{\sigma}$ for $\sigma$ away from zero.
Still more precisely and explicitly:

\begin{Thm}                      \label{Thm:Hypergeometric}
Let $P$ be a binomial, hypergeometric, or Poisson law,
with mean $\mu$ and standard deviation $\sigma>0$.
Then
\begin{equation}     \label{Eq:hypergeometric_CLT_error_bound}
 \left| P(I) - \left( \Phi\big(\text{\footnotesize$\frac{b-\mu}{\sigma}$}\big)
           - \Phi\big(\text{\footnotesize$\frac{a-\mu}{\sigma}$}\big) \right) \right|
 \ \le \ \frac{0.6879}{\sigma}
\end{equation}
holds for {\em every} intervall $I \subseteq \R$ with boundary points
$a,b\in[-\infty,\infty]$ with $a\le b$,
and
\begin{equation}                  \label{Eq:lower_concentration_variance_bound}
  \text{\rm L.H.S.\eqref{Eq:hypergeometric_CLT_error_bound}}
 \ \ge \ \frac{1}{\sqrt{1+12\sigma^2}}
 \ \ge \ \frac{0.2773}{1\vee\sigma}
\end{equation}
holds for some such interval.
\end{Thm}

\begin{Remarks}                           \label{Rems:Hypergeometric}
We consider the situation and notation of Theorem~\ref{Thm:Hypergeometric}.
\begin{parts}
\item  \label{part:Hyp_I_unbounded}               
 If $I$ is unbounded, that is, being one
 of $\,]-\infty,b[$\,, $]-\infty,b]\,$, $]a,\infty[$\,, $[a,\infty[\,$,
 then the upper bound in~\eqref{Eq:hypergeometric_CLT_error_bound}
 can be halved,
 and the lower bounds in~\eqref{Eq:lower_concentration_variance_bound}
 should be halved.
\item                     \label{part:Hyp_distr_fcts}
 We may rephrase part~\ref{part:Hyp_I_unbounded}
 and simultaneously introduce lower and upper approximation error bounds
 of the same simple form $\frac{c}{1\vee\sigma}$ as follows:
 With $F$ being the distribution function of~$P$ and
 $G(x)\coloneqq\Phi(\frac{x-\mu}{\sigma})$ for $x\in\R$,
 we have
 \begin{equation} \label{Eq:hypergeometric_CLT_error_bound_distr_fct}
   \frac{0.1386}{1\vee\sigma}
    \ \le \ \frac{1}{2\sqrt{1+12\sigma^2}} \ \le\ \left\|F-G\right\|_\infty
    \ \le \ 0.5410\wedge\frac{0.3440}{\sigma}
    \ \le \ \frac{0.5410}{1\vee \sigma} \,.
 \end{equation}
\item                                 \label{part:Hyp_distr_two-sided}   
 In analogy to~\eqref{Eq:hypergeometric_CLT_error_bound_distr_fct},
 we can rewrite~\eqref{Eq:hypergeometric_CLT_error_bound}
 and~\eqref{Eq:lower_concentration_variance_bound} as
 \begin{equation} \label{Eq:hypergeometric_CLT_error_bound_distr_two-sided}
   \frac{0.2773}{1\vee\sigma}
    \ \le \ \frac{1}{\sqrt{1+12\sigma^2}}
    \ \le\ \max_{I}\text{\rm L.H.S.\eqref{Eq:hypergeometric_CLT_error_bound}}
    \ \le \ 1\wedge \frac{0.6879}{\sigma}
    \ \le \ \frac{1}{1\vee \sigma}
 \end{equation}
 with the maximum taken over all intervals~$I$ as in Theorem~\ref{Thm:Hypergeometric}.
\item                        \label{part:Practically_interpretable_CLT}  
 With $\widetilde{F}$ denoting the standardisation of $F$
 from part~\ref{part:Hyp_distr_fcts},
 we have $ \left\|F-G\right\|_\infty =  \| \widetilde{F}-\Phi\|^{}_\infty\,$,
 and hence the upper bounding for $ \left\|F-G\right\|_\infty$
 in~\eqref{Eq:hypergeometric_CLT_error_bound_distr_fct} yields
 by specialisation to the binomial case, $P=\mathrm{B}_{n,p}$
 with $\mu=np$ and $\sigma=\sqrt{np(1-p)}$, a quantitative and hence practically
 interpretable version of~\eqref{Eq:binomial_CLT}.
\item                  \label{part:Bernoulli_convolutions}       
 The inequalities
  \eqref{Eq:hypergeometric_CLT_error_bound},
  \eqref{Eq:lower_concentration_variance_bound},
  \eqref{Eq:hypergeometric_CLT_error_bound_distr_fct},
  \eqref{Eq:hypergeometric_CLT_error_bound_distr_two-sided}
 hold more generally than stated above:

 We call any law of the form $\bigconv_{j=1}^n\mathrm{B}_{p_j}$
 with $p\in[0,1]^n$ a \emph{Bernoulli convolution}
 (also known as a \emph{Poisson binomial law}),
 write  $\check{P}$ for the reflection of any law $P$,
 defined by $\check{P}(B)\coloneqq P(-B)$ for $B\subseteq\R$ measurable,
 and we let here $\cP$ denote the set of all
 limit laws (say with respect to weak convergence, but here
 actually then equivalently with respect
 to convergence in total variation distance
 and convergence of all moments)
 of  convolutions $P\ast\check{Q}$ with $P,Q$ Bernoulli convolutions.
 Then $\cP$ contains all the laws $P$ considered in Theorem~\ref{Thm:Hypergeometric},
 and the inequalities
 \eqref{Eq:hypergeometric_CLT_error_bound},
  \eqref{Eq:lower_concentration_variance_bound},
  \eqref{Eq:hypergeometric_CLT_error_bound_distr_fct},
  \eqref{Eq:hypergeometric_CLT_error_bound_distr_two-sided}
 hold for every $P\in\cP$ with $\sigma(P)>0$.

 The lower boundings for the approximation errors in
 \eqref{Eq:lower_concentration_variance_bound},
  \eqref{Eq:hypergeometric_CLT_error_bound_distr_fct},
  \eqref{Eq:hypergeometric_CLT_error_bound_distr_two-sided}
 even hold if $P$ is any law on $\Z$ with mean $\mu$ and
 standard deviation $\sigma\in\mathopen]0,\infty\mathclose[\,$.
\item                                   \label{part:Constants_not_optimal} 
 The constants $0.6879$ in~\eqref{Eq:hypergeometric_CLT_error_bound}
 and $0.3440$ in~\eqref{Eq:hypergeometric_CLT_error_bound_distr_fct}
 are most likely not optimal,
 not even for the more general Bernoulli convolution case of part~\ref{part:Bernoulli_convolutions}.
 They  can be improved in the binomial or Poisson case as indicated in
 section~\ref{sec:Binomial} below,
 but, even for these special cases,
 $0.6879$ in~\eqref{Eq:hypergeometric_CLT_error_bound}
 can not in general be decreased beyond
 $\frac{1}{\sqrt{2\mathrm{e}}}=0.4288\ldots>\frac{1}{\sqrt{2\pi}}$,
 and $0.3440$ in~\eqref{Eq:hypergeometric_CLT_error_bound_distr_fct}
 can not in general be decreased beyond
 $0.2660>\frac{2}{3\sqrt{2\pi}}> \frac{1}{2\sqrt{2\mathrm{e}}}$\,,
 although, based on certain asymptotics,
 $\frac{1}{\sqrt{2\pi}}$ and $\frac{2}{3\sqrt{2\pi}}$
 would be quite natural conjectures for the
 optimal constants.
 In the rather special case of symmetric hypergeometric or binomial
 laws, $0.6879$ can be replaced with
 $\frac{1}{\sqrt{2\pi}}=0.3989\ldots\,$,
 and $0.3440$ with $\frac{1}{2\sqrt{2\pi}}=0.1994\ldots\,$.
\item                       \label{part:Equality_in_lower_bound}
 In the first inequality in~\eqref{Eq:lower_concentration_variance_bound},
 we have equality for $P=\mathrm{B}_{1,\frac{1}{2}}$
 with then $\sigma^2=\frac{1}{4}$
 (considering $I=\{0\}$ or $I=\{1\}$),
 and also in the limit $\sigma\rightarrow0$ with otherwise $P$ arbitrary
 (considering $I=\{k\}$ with $k$ the integer nearest to $\mu$).
 But for the only interesting case of large $\sigma$,
 some improvement should be possible.
\item                           \label{part:WR}
 The present
 proof of the upper bounds for the approximation error in Theorem~\ref{Thm:Hypergeometric}
 uses  \citeposs{Shevtsova2013} theorem stated as \eqref{Eq:Shevtsova_2013_i.a.d.} below,
 which has a complicated proof, combining Fourier analytic methods with
 computerised numerical calculations.
 With just some universal constant in place of $0.6879$,
 Theorem~\ref{Thm:Hypergeometric} is proved along classical lines (using just Stirling's formula,
 Taylor approximations, and the Euler-MacLaurin summation formula)
 in \citet[pp.~126--138]{MattnerWR}.
\end{parts}
\end{Remarks}

\section{Normal approximations to binomial interval probabilities,
 and hence also to Poisson ones}      \label{sec:Binomial}

In the binomial case, and hence also in the limiting Poisson case,
one can sharpen Theorem~\ref{Thm:Hypergeometric} a bit. Since the pure binomial
case might be the first or even the only normal approximation result some students will see,
let us state it here separately, without reference to Poisson laws:

\begin{Thm}                          \label{Thm:Binomial}
Let $P$ be a binomial law with the parameters $n\in\N$ and $p\in\mathrm]0,1\mathclose[\,$,
and hence with mean $\mu=np$ and standard deviation $\sigma=\sqrt{np(1-p)}>0$.
Then
\begin{equation}     \label{Eq:binomial_CLT_error_bound}
 \left| P(I) - \left( \Phi\big(\text{\footnotesize$\frac{b-\mu}{\sigma}$}\big)
           - \Phi\big(\text{\footnotesize$\frac{a-\mu}{\sigma}$}\big) \right) \right|
 \ \le \ \frac{0.6379}{\sigma}
\end{equation}
holds for {\em every} intervall $I \subseteq \R$ with boundary points
$a,b\in[-\infty,\infty]$ with $a\le b$, and
\begin{equation}                  \label{Eq:lower_concentration_variance_bound_repeated}
  \text{\rm L.H.S.\eqref{Eq:binomial_CLT_error_bound}}
 \ \ge \ \frac{1}{\sqrt{1+12\sigma^2}}
 \ \ge \ \frac{0.2773}{1\vee\sigma}
\end{equation}
holds for some such interval.
\end{Thm}

\begin{Remarks}                           \label{Rems:Binomial}
\begin{parts}
\item                                      \label{part:Binomial_is_analogous}
 Remarks~\ref{Rems:Hypergeometric}, with $0.6879$ changed to $0.6379$
 and $0.3440$ changed to $0.3190$, apply with the exception
 of~\ref{Rems:Hypergeometric}\ref{part:Bernoulli_convolutions}.
 Thus the analogues of~\eqref{Eq:hypergeometric_CLT_error_bound_distr_fct}
 and~\eqref{Eq:hypergeometric_CLT_error_bound_distr_two-sided}
 are
 \begin{equation} \label{Eq:Binomial_CLT_error_bound_distr_fct}
   \frac{0.1386}{1\vee\sigma}
    \ \le \ \frac{1}{2\sqrt{1+12\sigma^2}} \ \le\ \left\|F-G\right\|_\infty
    \ \le \ 0.5410\wedge\frac{0.3190}{\sigma}
    \ \le \ \frac{0.5410}{1\vee \sigma}
 \end{equation}
 and
 \begin{equation} \label{Eq:Binomial_CLT_error_bound_distr_two-sided}
   \frac{0.2773}{1\vee\sigma}
    \ \le \ \frac{1}{\sqrt{1+12\sigma^2}}
    \ \le\ \max_{I}\text{\rm L.H.S.\eqref{Eq:binomial_CLT_error_bound}}
    \ \le \ 1\wedge \frac{0.6379}{\sigma}
    \ \le \ \frac{1}{1\vee \sigma}\,.
 \end{equation}
\item Better upper bounds for $\left\|F-G\right\|_\infty$
  and $\max_{I}\text{\rm L.H.S.\eqref{Eq:binomial_CLT_error_bound}}$,
  not insisting on the form $c/\sigma = c/\sqrt{np(1-p)}$
  but allowing more generally $c(p)/\sigma = \big(c(p)/\sqrt{p(1-p)}\,\big)/\sqrt{n}$,
  can be given for $p$ not too far away from $\frac{1}{2}$, namely
  \begin{align}                \label{Eq:Bin_p_close_to_1/2}
   \left\|F-G\right\|_\infty \ \le \ \frac{\frac{1}{2}+\frac{1}{3}|p-\frac{1}{2}|}{\sqrt{2\pi}}\cdot \frac{1}{\sigma}
        \ \le \ \frac{5}{9\sqrt{2\pi}\,\sigma}
        \ = \ \frac{0.2216\ldots}{\sigma} \quad\text{ if } p\in[\tfrac{1}{3},\tfrac{2}{3}]
  \end{align}
  with asymptotic equality for $n\rightarrow\infty$ in the first inequality, and
  \begin{align}              \label{Eq:Bin_BE_Jona}
   \left\|F-G\right\|_\infty \ \le \ \frac{3+\sqrt{10}}{6\sqrt{2\pi}}
    \cdot \frac{ \frac{1}{2} + 2\,\big(p-\frac{1}{2}\big)^2}{\sigma}   \quad\text{ if } p\in\mathopen]0,1\mathclose[
  \end{align}
  with $\text{R.H.S.}< \frac{0.3190}{\sigma}$ iff $|p-\frac{1}{2}| <0.3732\ldots\,$.
\item For a Poisson law with parameter $\lambda>0$, we have to put in the  above
 $\mu\coloneqq \lambda$, $\sigma\coloneqq\sqrt{\lambda}$, and $p=0$;
 so \eqref{Eq:Bin_p_close_to_1/2} does not apply,
 and \eqref{Eq:Bin_BE_Jona} is worse than~\eqref{Eq:Binomial_CLT_error_bound_distr_fct}.
\end{parts}
\end{Remarks}

%

\section{Refined normal approximations to symmetric binomial laws:
 the benefit of the continuity correction} \label{sec:Symmetric_binomial}
The special case of symmetric binomial laws
is of interest due to its usefulness in statistics
for performing a quick sign test,
and also since a simple sharpening of Theorem~\ref{Thm:Binomial}
holds:

\begin{Thm}                          \label{Thm:Symmetric_binomial}
Let $P$ be a symmetric binomial law with the parameter $n\in\N$,
and hence with mean $\mu=\frac{n}{2}$ and variance $\sigma^2=\frac{n}{4}$.
Then
\begin{equation}     \label{Eq:Symmetric_binomial_CLT_error_bound_with_continuity_correction}
 \left| P(I) - \left( \Phi\big(\text{\footnotesize$\frac{b-\mu}{\sigma}$}\big)
           - \Phi\big(\text{\footnotesize$\frac{a-\mu}{\sigma}$}\big) \right) \right|
 \ \le \ \frac{\Phi(-3/\sqrt{2}\,)}{\sigma^2}
 \ =\  \frac{0.01694\ldots}{\sigma^2}
 \ = \ \frac{0.06778\ldots}{n}
\end{equation}
holds for every interval $I\subseteq\R$ with boundary points
$a,b\in \big\{k+\frac{1}{2} : k\in\Z\big\} \cup \big\{-\infty,\infty\big\}$
with $a\le b$.
\end{Thm}

\begin{Remarks}                           \label{Rems:Symmetric_binomial}
We consider the situation and notation of Theorem~\ref{Thm:Symmetric_binomial}.

\begin{parts}
\item To approximate $P(\{k,\ldots,\ell\})$ for any $k,\ell\in\Z$ with $k\le\ell$,
 one applies Theorem~\ref{Thm:Symmetric_binomial} with $a\coloneqq k-\frac{1}{2}$
 and $b\coloneqq\ell+\frac{1}{2}\,$.
\item                         \label{part:Symmetric_binomial_CLT_error_bound_with_CC_one-sided}
 If $I$ is unbounded, then the error bound
 in~\eqref{Eq:Symmetric_binomial_CLT_error_bound_with_continuity_correction}
 can be halved.
\item The error bounds
 in~\eqref{Eq:Symmetric_binomial_CLT_error_bound_with_continuity_correction}
 and in \ref{part:Symmetric_binomial_CLT_error_bound_with_CC_one-sided}
 are of the correct order in $n$, with in each of the two cases equality occurring iff $n=2$
 and $I$ is maximal with $P(I)=1$.
\item There is no simple analogue of Theorem~\ref{Thm:Symmetric_binomial}
 for general binomial laws.
\end{parts}
\end{Remarks}

\section{Bounds for the Clopper-Pearson confidence bounds}  \label{sec:Bounds_for_CP_bounds}
\newcommand{\pu}{\underline{p}{}^{}}
\newcommand{\puu}{\underline{\underline{p}}{}^{}}
\newcommand{\puuu}{\underline{\underline{\underline{p}}}{}^{}}
\newcommand{\puuuu}{{\underline{\underline{\underline{\underline{p}}}}}{}^{}}
\newcommand{\po}{\overline{p}{}^{}}
\newcommand{\poo}{\overline{\overline{p}}{}^{}}
\newcommand{\pooo}{\overline{\overline{\overline{p}}}{}^{}}
\newcommand{\poooo}{{\overline{\overline{\overline{\overline{p}}}}}{}^{}}
%
%
The arguably most basic problem of statistical inference is to estimate,
with a confidence interval, an unknown success probability parameter
from
a binomially distributed observable with known sample size parameter $n$.
Given a confidence level $\beta\in\mathopen]0,1\mathclose[$\,, for example $\beta=0.95$,
and considering first only one-sided intervals,
expressible through lower or upper confidence bound $\pu_\beta=\pu_{n,\beta}$
and $\po_\beta=\po_{n,\beta}$\,,
the problem is solved optimally by \citeposs{ClopperPearson1934}
\begin{align}                  \label{Eq:Def_pu}
 \pu_\beta(x) &\,\coloneqq\, \left\{ \begin{array}{ll} 0&\quad\text{ if }x=0,\\
   \text{the }p\in[0,1]\text{ with }\mathrm{B}_{n,p}(\{x,\ldots,n\})=1-\beta
   &\quad\text{ if }x>0 \,,
 \end{array}\right.  \\
  \po_\beta(x)  &\,\coloneqq \, 1-\pu_\beta(n-x)   
                                                   \label{Eq:Def_po}
\end{align}
for $x\in\{0,\ldots,n\}$.
Simple explicit formulae for these exist apparently in the following cases only:
\begin{align}
 &\pu_\beta(0)  \,=\,0\,, \qquad
  \pu_\beta(1)  \,=\,1-\beta^{1/n}\,,  \qquad
  \pu_\beta(n)  \,=\,(1-\beta)^{1/n}\,,                      \label{Eq:p_l_beta(1),p_beta(n)} \\
 &\po_\beta(0) \,=\,1-(1-\beta)^{1/n}\,, \qquad
  \po_\beta(n-1) \,=\, \beta^{1/n}\,, \qquad                   
   \po_\beta(n) \,=\,         1 \,.           \label{Eq:Bin.conf.boundary}
\end{align}

In any serious application of~\eqref{Eq:Def_pu} or~\eqref{Eq:Def_po}
with $\beta, n,x$ given,
one will, at least if~\eqref{Eq:p_l_beta(1),p_beta(n)}
or~\eqref{Eq:Bin.conf.boundary} does not apply,
use some appropriate software, for example~R, and this should of course be taught.
But for a rough orientation or even a back-of-the-envelope calculation,
or perhaps for planning a suitable sample size $n$ if possible,
one wants simple, if somewhat rough, approximations to the Clopper-Pearson
bounds. The following theorem bounds each Clopper-Pearson bound in the more interesting
direction.

\begin{Thm}                                       \label{Thm:CP-bounds_bounded}
Let $n\in \N$ and $\beta\in[\frac{1}{2},1[$\,.
Then the lower and upper Clopper-Pearson (\citeyear{ClopperPearson1934}) $\beta$-confidence bounds
$\underline{p} = \underline{p}_{\,n,\beta}$
and $\overline{p} = \overline{p}_{n,\beta}$
for the binomal success probability parameter
admit the boundings
\begin{align}
  \underline{p}(x) \ &>\ \widehat{p}
    \,-\, \frac{z}{\sqrt{n}}\sqrt{\widehat{p}\widehat{q} +\frac{1+\frac{z^2}{9}}{n}}
    \,-\, \frac{1+\frac{z^2}{3}}{n}\, , \label{Eq:lower_CP-bound_nicely_bounded}  \\
  \overline{p}(x) \ &<\ \widehat{p}
    \,+\, \frac{z}{\sqrt{n}}\sqrt{\widehat{p}\widehat{q} +\frac{1+\frac{z^2}{9}}{n}}
    \,+\, \frac{1+\frac{z^2}{3}}{n} \label{Eq:upper_CP-bound_nicely_bounded}
\end{align}
for $x\in\{0,\ldots,n\}$, where $\widehat{p}\coloneqq\frac{x}{n}$,
$\widehat{q}\coloneqq 1-\widehat{p}$, and $z\coloneqq\Phi^{-1}(\beta)$.
\end{Thm}

\begin{Remarks}                           \label{Rems:CP-bounds_bounded}
In the situation and with the notation of Theorem~\ref{Thm:CP-bounds_bounded},
let further $\cX\coloneqq\{0,\ldots,n\}$ and, for $x\in\cX$,
\begin{align}                        \label{Eq:Def_simple_bounds_for_CP_in_sec_4}
 \underline{\underline{p}}(x) \,\coloneqq\,\text{\rm R.H.S.\eqref{Eq:lower_CP-bound_nicely_bounded}}\,,\qquad
 \overline{\overline{p}}(x) \,\coloneqq\,\text{\rm R.H.S.\eqref{Eq:upper_CP-bound_nicely_bounded}}\,.
\end{align}
\begin{parts}
\item For two-sided $\beta$-confidence intervals one may (or even should?) combine
 $\underline{p}_{\,n,\frac{1+\beta}{2}}$ with $\overline{p}_{n,\frac{1+\beta}{2}}$.
\item
 We have the symmetry $\overline{p}(x) = 1-\underline{p}(n-x)$ by definition,
 but obviously also
$\overline{\overline{p}}(x) = 1-\underline{\underline{p}}(n-x)$,
 for $x\in\cX$.
 Hence it suffices in what follows to consider, say, only upper confidence bounds.
 For example, taken together, the inequalities in~\eqref{Eq:lower_CP-bound_nicely_bounded}
 are equivalent to the ones in~\eqref{Eq:upper_CP-bound_nicely_bounded}.

\item The simple \citet{Lagrange1776} formula
 $\widetilde{p} \coloneqq \widehat{p} +\frac{z}{\sqrt{n}}\sqrt{\widehat{p}\widehat{q}}$
 is well-known to be invalid as an upper $\beta$-confidence bound even asymptotically.
\item
 But $\widetilde{p}$ approximates $\overline{p}$ up to $O(1/n)$ for
 $\widehat{p}$ 
 bounded away from $0$ and $1$,
 so the valid bound $\overline{\overline{p}}$ is a good substitute.
\item In the example of $x=0$, we have
\begin{align}
 \overline{p}(0) & \ = \ 1-(1-\beta)^\frac{1}{n} \ < \ \frac{-\log(1-\beta)}{n} \text{ always},\quad
   \ \approx \text{and} < \,\ \frac{3}{n}\,\ \text{ if } \beta=0.95\,,    \\
 \widetilde{p}(0) & \ = \ 0\,, \\
 \overline{\overline{p}}(0) & \ = \ \frac{z\sqrt{1+\frac{z^2}{9}}+1+\frac{z^2}{3}}{n} \text{ always},\quad
   \ = \ \frac{3.777\ldots}{n} \,\ \text{ if } \beta=0.95\,.
\end{align}
\end{parts}
\end{Remarks}

\section{Proofs for sections~\ref{sec:Hypergeometric}\,--\,\ref{sec:Symmetric_binomial}}   \label{sec:Proofs_1--3}
\begin{proof}[Proofs for Theorem~\ref{Thm:Hypergeometric}
              and Remarks~\ref{Rems:Hypergeometric}] 

We recall that for any law $P$ on $\R$,  we let $F$ denote its distribution function
and, if $\sigma(P)\in\mathopen]0,\infty\mathclose[\,$,
write $\widetilde{F}$ for its standardised distribution function.

\smallskip 1.
 Let $\cP$ be defined as in
 Remark~\ref{Rems:Hypergeometric}\ref{part:Bernoulli_convolutions}.
 \citet{VatutinMikhailov1982} showed that every
 hypergeometric law is a Bernoulli convolution.
 \pr{They used this to obtain
 an upper normal approximation error bound like~\eqref{Eq:Esseen-Shevtsova_for_cP},
 as is also explained by~\citet[p.~733]{MattnerSchulz2018}.
 The latter were able to write down a numerically sharper
 version than \citet{VatutinMikhailov1982}
 by using \citeposs{Shevtsova2013} result~\eqref{Eq:Shevtsova_2013_i.a.d.}, but they overlooked
 the still better version~\eqref{Eq:Esseen-Shevtsova_for_cP} given below.}
 Since $\cP$ trivially contains all binomial laws,
 and hence all Poisson laws as limits of binomial laws,
 the claim of the second paragraph
 of Remark~\ref{Rems:Hypergeometric}\ref{part:Bernoulli_convolutions}
 is now obvious, except perhaps for the claim that
 weak convergence of laws in $\cP$ implies convergence
 of all moments, which is true by
 \citet[Satz 4.10 on p.~27, Bemerkung 5.10 und Lemma 5.11
 on p.~37]{Tasto2011}.

\smallskip 2.
Let $\Prob_3(\R)$ denote the set of all laws on $\R$ with finite third absolute moments.
According to \citet{Shevtsova2013}, we have the following refinement
of \citeposs{Esseen1945} theorem:

\emph{Let $n\in\N$ and $P_1,\ldots,P_n\in\Prob_3(\R)$ with the
 corresponding standard deviations $\sigma_j$
 and third centred absolute moments
 $\beta_j\coloneqq\int |x-\mu(P_j)|^3\,\dd P_j(x)$,
 and with $\sigma\coloneqq\sqrt{\sum_{j=1}^n\sigma_j^2}>0$\,.
 Let $L\coloneqq \sigma^{-3/2}\sum_{j=1}^n\beta_j$ (the Lyapunov ratio)
 and $T \coloneqq \sigma^{-3/2}\sum_{j=1}^n \sigma_j^3$.
 Then the standardised distribution function $\widetilde{F}$
 of the convolution $\bigconv_{j=1}^n P_j$ satisfies
 \begin{equation}             \label{Eq:Shevtsova_2013_i.a.d.}
   \left\| \widetilde{F}-\Phi\right\|_\infty \ \le \
    \min\left\{\,0.5583\,L\,,\
                 0.3723\,\big(L+\tfrac{1}{2}T\big)\,,\
                 0.3057\,\big(L+T\big)
     \,\right\}.
 \end{equation}
 }

 \pr{\citet{Shevtsova2013} assumes for notational convenience
  $\mu(P_j)=0$ and $\sigma=1$.
  This yields the present version via
  random variables
  $X_j\coloneqq \frac{1}{\sigma}(Y_j-\mu(P_j))$ with independent $Y_j\sim P_j$.

  The interest of the additonal terms beyond $0.5583\,L$ in the minimum in~\eqref{Eq:Shevtsova_2013_i.a.d.}
  is due to $\sigma_j^3\le \beta_j$  and hence $T\le L$ always,
  but possibly $T$ much smaller than $L$.}

 In the special case of Bernoulli laws, $P_j=\mathrm{B}_{p_j}$
 with $p\in[0,1]^n$, we have
 $\sigma_j = \sqrt{p_j(1-p_j)}$
 and $\beta_j=(p_j^2+(1-p_j)^2)p_j(1-p_j)=(1-2\sigma_j^2)\sigma_j^2$,
 and with $x\coloneqq \sqrt{\sigma^{-2}\sum_{j=1}^n\sigma_j^4}$
 hence \eqref{Eq:Shevtsova_2013_i.a.d.} yields
 \begin{align}                   \label{Eq:Esseen-Shevtsova_for_BCs}
  \sigma\left\|\widetilde{F}-\Phi\right\|_\infty
   &\ \le\  0.3057\,\big(\sigma L+\sigma T\big) \\ \nonumber
   &\  = \ 0.3057\,\big( 1- 2\sigma^{-2}\sum_{j=1}^n\sigma_j^4
            + \sigma^{-2}\sum_{j=1}^n \sqrt{\sigma_j^2} \sigma_j^2 \big)  \\ \nonumber
   &\ \le\  0.3057\,\big( 1- 2x^2 +x\big)
    \ \le\ 0.3057\cdot\frac{9}{8}
    \ = \ 0.3439125
 \end{align}
 by using in the third step Jensen's inequality with the
 concave function $\sqrt{\cdot}$,
 and by maximising in the fourth step the quadratic function at $x=\frac{1}{4}$.

 Since reflected Bernoulli convolutions are shifted Bernoulli convolutions,
 and by the remark about convergence of moments,
 already justified in part 1 of this proof and yielding convergence of the mean and the standard deviation,
 we see that \eqref{Eq:Esseen-Shevtsova_for_BCs} implies
 \begin{align}                   \label{Eq:Esseen-Shevtsova_for_cP}
  \left\|\widetilde{F}-\Phi\right\|_\infty
   &\ \le\ \frac{ 0.3439125}{\sigma(P)} \quad\text{ for $P\in\cP$ with $\sigma(P) >0$}\,.
 \end{align}

\smallskip 3.
 If $P$ be any law on $\Z$ with $\sigma\coloneqq\sigma(P)<\infty$,
 then $\max_{x\in\Z}P(\{x\})\ge(1+12\sigma^2)^{-1/2}$.
 This follows from an inequality of \citet{Levy1937},
 namely \citet[p.~149, Lemme 48.1]{Levy1954},
 or more directly from a weaker but simpler consequence of
 L\'evy's inequality obtained apparently independently
 by \citet[p.~981, Corollary 2.2]{BobkovChistyakov2015}.
 See for example
 \citet[p.~733, (18) with $h=1$]{MattnerSchulz2018}
 for a short proof and further references.

 We have $(1+12\sigma^2)^{-1/2} \ge 13^{-1/2}(1\vee\sigma)^{-1}$,
 $ 13^{-1/2} = 0.277350\ldots\;$.

\smallskip 4.
 If $F$ and $G$ are distribution functions with identical means and variances,
 and with $G$ normal, then $\left\| F-G\right\|_\infty<  0.5409365\ldots$
 holds by \citeposs{ChebotarevEtAl2007} refinement of
 \citet[p.~103, Lemma 12.3]{BhatRangaRao2010}.

\smallskip 5. Combining the above parts 1--4 easily yields
 Remark~\ref{Rems:Hypergeometric}\ref{part:Bernoulli_convolutions},
 and hence Theorem~\ref{Thm:Hypergeometric} and
 Remarks~\ref{Rems:Hypergeometric}\ref{part:Hyp_I_unbounded},
 \ref{part:Hyp_distr_fcts},
 \ref{part:Hyp_distr_two-sided},
 \ref{part:Practically_interpretable_CLT}.

\smallskip 6. In Remark~\ref{Rems:Hypergeometric}\ref{part:Constants_not_optimal},
the lower bound $\frac{1}{\sqrt{2\mathrm{e}}}$ for the optimal
constant possibly replacing $0.6879$
in~\eqref{Eq:hypergeometric_CLT_error_bound}
is obtained by considering the Poisson law with parameter $\lambda=\frac{1}{2}$
and the degenerate interval $I=\{0\}$,
as suggested by a result of \citet{Herzog1947}.
The lower bound $0.2660$ is obtained in
\citet[p.~16,
where $C_{\mathrm{BE},\{ \mathrm{Poi}_\alpha :\alpha \in (0,\infty)\}}
> \frac{2}{3\sqrt{2\pi}} > 0.2660\ldots$
should be read as
$C_{\mathrm{BE},\{ \mathrm{Poi}_\alpha :\alpha \in (0,\infty)\}}
 \ge 1.00018989 \frac{2}{3\sqrt{2\pi}} = 0.2660\ldots$]
{Schulz2016}.
The asymptotics alluded to in connection with $\frac{1}{\sqrt{2\pi}}$ and $\frac{2}{3\sqrt{2\pi}}$
are specialisations to the binomial case
of results of \citet{DinevMattner2012} and \citet{Esseen1956},
as given in \citet[Remark 4.4 on p.~17, Remark 4.2 on p.~15]{Schulz2016}.
The claim about the symmetric case follows from \citet[pp.~731--732, (12), (14)]{MattnerSchulz2018},
which is proved there using direct computations with densities;
here the simpler binomial subcase was obtained earlier by \citet[Corollary~1.2]{HippMattner2007},
and later differently, using the characteristic function method,
by \citet[p.~16, Theorem VIII]{vanNerven2024}.

\smallskip7.
Remarks~\ref{Rems:Hypergeometric}\ref{part:Equality_in_lower_bound}
and \ref{part:WR} are easily checked.
\end{proof}

\begin{proof}[Proofs for Theorem~\ref{Thm:Binomial} and Remarks~\ref{Rems:Binomial}]
\mbox{}

\smallskip1. Again according to \citet{Shevtsova2013}, inequality \eqref{Eq:Shevtsova_2013_i.a.d.}
 can in the case of identical $P_1=\ldots=P_n$ be sharpened to
 \begin{equation}             \label{Eq:Shevtsova_2013_i.i.d.}
   \left\| \widetilde{F}-\Phi\right\|_\infty \ \le \
    \min\left\{\,0.4690\,L\,,\
                 0.3322\,\big(L+0.429\,T\big)\,,\
                 0.3031\,\big(L+0.646\,T\big)
     \,\right\}.
 \end{equation}
 Hence for a binomial law $P=\mathrm{B}_{n,p}$ with standardised distribution function $\widetilde{F}$,
 we get instead of~\eqref{Eq:Esseen-Shevtsova_for_BCs} now, with $x\coloneqq \sqrt{p(1-p)}$,
 \begin{align}                   \label{Eq:Esseen-Shevtsova_for_Bin}
  \sigma\left\|\widetilde{F}-\Phi\right\|_\infty
   &\ \le\  0.3031\,\big(\sigma L+ 0.646\,\sigma T\big) \\ \nonumber
   &\  = \ 0.3031\,\big( 1- 2 x^2
            + 0.646\, x \big)
    \ \le\  0.31891105995
 \end{align}
 by maximising the quadratic function at $x=0.1615\,$.

 This easily implies Theorem~\ref{Thm:Binomial} and Remark~\ref{Rems:Binomial}\ref{part:Binomial_is_analogous}.

\smallskip2. \eqref{Eq:Bin_p_close_to_1/2} and \eqref{Eq:Bin_BE_Jona}
are the main results of \citet{Schulz2016}.
\end{proof}

\begin{proof}[Proofs for Theorem~\ref{Thm:Symmetric_binomial} and
 Remarks~\ref{Rems:Symmetric_binomial}]
 See \citet[Theorem IV on p.~13, discussion of classical results
  for general binomial laws on p.~6, Theorem XII on p.~19]{vanNerven2024}.
\end{proof}

\section{Proofs for and complements to section \ref{sec:Bounds_for_CP_bounds}}  \label{sec:Proofs_4}

In this section, we change the notation~\eqref{Eq:Def_simple_bounds_for_CP_in_sec_4}
used in Remarks~\ref{Rems:CP-bounds_bounded} to
\begin{align}                        \label{Eq:Def_simple_bounds_for_CP_in_sec_6}
 \puuuu_\beta(x) \,\coloneqq\,\text{\rm R.H.S.\eqref{Eq:lower_CP-bound_nicely_bounded}}\,,\qquad
 \pooo_\beta(x) \,\coloneqq\,\text{\rm R.H.S.\eqref{Eq:upper_CP-bound_nicely_bounded}}\,.
\end{align}
Theorem~\ref{Thm:CP-bounds_bounded} is contained in the following more refined result.

\begin{Thm}                    \label{Thm:First_bounds_for_Clopper-Pearson_bounds}
Let $n\in\N$, $x\in\{0,\ldots,n\}$, $\hat{p}\coloneqq\frac{x}{n}$, $\hat{q}\coloneqq1-\hat{p}$,
$\beta\in[\frac12,1[\,$, and $z\coloneqq \Phi^{-1}(\beta)$. Then
the Clopper-Pearson bounds $\pu_\beta(x)$ and $\po_\beta(x)$
from~\eqref{Eq:Def_pu} and~\eqref{Eq:Def_po} satisfy the following boundings:

\smallskip
If $x\ge 1$, then with $u\coloneqq \widehat{p}-\frac{1}{n}$
\la  \label{Eq:Clopper-Pearson_lower_bound_lower_bounded}
 \quad
 \pu_\beta(x) & > & \puu_\beta(x)
   \,\ \coloneqq \,\
     u - \frac{1}{1+\tfrac{z^2}{3n}}\Big(  \tfrac{z^2(2u-1)}{3n}
     + \frac{z}{\sqrt{n}} \sqrt{ u(1-u) + \tfrac{z^2(1-u(1-u))}{9n}  }\  \Big) \\ \nonumber
  &\ge&  \puuu_\beta(x)
    \,\ \coloneqq\,\
     \hat{p} -
    \frac{1}{1+\tfrac{z^2}{3n}}\Big( \frac{z}{\sqrt{n}}\sqrt{ \hat{p}\hat{q}
       + \tfrac{1+\frac{z^2}{9}}{n}} + \tfrac{1+\frac{z^2}{3}}{n}\,  \Big)\,.
\al

If $x\le n-1$, then
\la
 \po_\beta(x)        \label{Eq:Clopper-Pearson_upper_bound_upper_bounded}
   & <& \poo_\beta(x)
      \,\ \coloneqq \,\  1- \puu_\beta(n-x) \\ \nonumber
   &\le& \pooo_\beta(x)
    \,\ \coloneqq\,\ \hat{p} +
   \frac{1}{1+\tfrac{z^2}{3n}}\Big( \frac{z}{\sqrt{n}}\sqrt{ \hat{p}\hat{q}
       + \tfrac{1+\frac{z^2}{9}}{n}} + \tfrac{1+\frac{z^2}{3}}{n}\,  \Big) \,.
\al

If $x\in\{0,\ldots,n\}$ is unrestricted, then we have
\begin{equation}                 \label{Eq:CP_bounds_unrestricted}
  \pu_\beta(x) \,\ge\, \puuu_\beta(x) \,\ge\, \puuuu_\beta(x)\,, \qquad
  \po_\beta(x) \,\le\, \pooo_\beta(x) \,\le\, \poooo_\beta(x)\,.
\end{equation}
\end{Thm}

\medskip
Our proof of Theorem~\ref{Thm:First_bounds_for_Clopper-Pearson_bounds},
starting on page~\pageref{page:proof_of_CP_bounds} below,
refines \citeposs{Short2023} approach of combining \citeposs{ZubkovSerov2012}
Theorem~\ref{Thm:Zubkov-Serov} with
Lemma~\ref{Lem:Bernoulli_rate_function_lower_bound}
known from \citet{Janson2016}.
We avoid the detour through \citeposs[p.~4185, (3)]{Short2023}
inequality for binomial quantile functions. 
Let us mention here
that \citeposs[p.~4188, Corollary~3]{Short2023} upper bound for what we
call $\po_\beta$ is not completely correct, since for, in his notation,
$n,k\rightarrow \infty$ with $n-k$ fixed, and $R$ fixed but large enough,
the main radicand in his bound eventually becomes strictly negative.

\begin{Thm}[\citet{ZubkovSerov2012} inequalities, here marginally extended]
                                                     \label{Thm:Zubkov-Serov}
Let $n\in\N$, $p\in\mathopen]0,1\mathclose[\,$,
$F_{n,p}$ the distribution function of the binomial law $\mathrm{B}_{n,p}$,
and
\la                         \label{Eq:Def_H_Zubokov_Serov}
 H(u,p) &=& \left\{\begin{array}{ll}
   u\log\tfrac{u}{p} + (1-u)\log\tfrac{1-u}{1-p} & \text{ for  }u \in\mathopen[0,1\mathclose]\,, \\
   \infty & \text{ for } u\in\R\setminus[0,1]\,,
  \end{array} \right.
\al
with in particular $H(0,p)=\log\frac{1}{1-p}$
and $H(1,p) = \log \frac{1}{p}$\,. Let finally
\[
 C_{n,p}(k) &\coloneqq& \Phi\Big(\!\sgn(\tfrac{k}{n}-p) \sqrt{2n H(\tfrac{k}{n},p)}\,\Big) \quad\text{ for }k\in\Z\,.
\]
Then
\la                        \label{Eq:Zubkov_Serov}
  C_{n,p}(k) &\le & F_{n,p}(k) \,\ \le \,\ C_{n,p}(k\!+\!1) \quad\text{ for }k\in \Z\,,
\al
with the $\begin{Bmatrix} \!\!\!\text{left}\ \\ \text{right} \end{Bmatrix}$ hand
inequality strict iff $k\in \begin{Bmatrix} \{0,\ldots,n\}\qquad\mbox{} \\ \{-1,\ldots,n\!-\!1\} \end{Bmatrix}$.
\end{Thm}
\pr{$H(\cdot,p)$ is called the {\em large deviation rate function} of the Bernouli law $\mathrm{B}_p$.
See \citet[p.7, Theorem 3.1]{Bahadur1971} for the theorem explaining the name; other
sources often have additional assumptions, for example~\citet[pp.~5,9]{denHollander2000}
or \citet[p.~540]{Kallenberg2002}.}

\begin{proof}
%
1.  
In view of
$F_{n,p}(k) = 1- F_{n,1-p}(n\!-\!k\!-\!1)$ and $C_{n,p}(k) = 1-C_{n,1-p}(n-k)$
for each $(p,k)\in\mathopen]0,1\mathclose[ \times \Z$,
it is enough to consider one of the two inequalities in~\eqref{Eq:Zubkov_Serov}.
Thus
\citet{ZubkovSerov2012} prove  $ F_{n,p}(k) < C_{n,p}(k\!+\!1) $
for $(p,k)\in \mathopen]0,1\mathclose[ \times\{0,\ldots,n\!-\!2\}$,
and hence $ C_{n,p}(k) < F_{n,p}(k) $ for $(p,k)\in \mathopen]0,1\mathclose[ \times\{1,\ldots,n\!-\!1\}$.

We trivially have $ C_{n,p}(n) < 1 =F_{n,p}(n) $,
and $ C_{n,p}(k) = F_{n,p}(k)\in\{0,1\}$ for $k\in  \Z  \setminus \{0,\ldots,n\}$.
Hence it only remains to prove $C_{n,p}(0) <F_{n,p}(0)$.

\smallskip 2. We recall the standard normal tail inequality
\la                          \label{Eq:Standard_normal_tail_inequality}
 \Phi(-z) &<& \frac{\phi(z)}{z} \quad \text{ for } z \in\mathopen]0,\infty\mathclose[\,.
\al
Putting for $p\in\mathopen]0,1\mathclose[$
\[
 g(p) \, \coloneqq\, \log\frac{F_{n,p}(0)}{C_{n,p}(0)}
  \, =\, \log\frac{(1-p)^n}{\Phi(-z(p))} \ \text{ with } \
   z(p)\,\coloneqq\, \sqrt{2n\log\frac{1}{1-p}} 
\]
we obtain $g(0+)= \log(2)>0$ and, using~\eqref{Eq:Standard_normal_tail_inequality},
\[
 g'(p) &=&-\frac{n}{1-p} -\frac{\phi(-z(p))}{\Phi(-z(p))}(-z'(p))
  \,\ =\,\ \frac{n}{1-p}\left(-1 + \frac{\phi(z(p))}{z(p)\Phi(-z(p))} \right) \,\ >\,\ 0
\]
and hence $F_{n,p}(0) > 2C_{n,p}(0) > C_{n,p}(0)$.
\end{proof}

The following simple rational lower bound for the transcendental function $H$
from~
\eqref{Eq:Def_H_Zubokov_Serov} was, under the restriction $u\ge p$,
stated and proved in a different parametrisation
by \citet[p.~8, ``Hence $f(x)\ge 0$ in this interval'']{Janson2016},
and quoted in essentially the present parametrisation by
\citet[p.~4, (23)]{Short2013} and \citet[p.~4185, (8)]{Short2023}.
As seen below, Janson's proof, with minor details added, works without the restriction.

\pr{\citet[p.\,1]{Janson2016} writes: ``This paper was written in 1994, but was never published because I had overlooked some existing papers containing some of the inequalities. Because of some recent interest in one of the inequalities,
which does not seem to be published anywhere else, it has now been lightly edited and made available here''.
He unfortunately does not state which inequalities were already published earlier, and which seemed to be new then.}

\begin{Lem}
                                               \label{Lem:Bernoulli_rate_function_lower_bound}%
Let $p\in\mathopen]0,1\mathclose[\,$. With $H$ defined
by~{\rm\ref{Thm:Zubkov-Serov}}\eqref{Eq:Def_H_Zubokov_Serov},
we then have
\la                              \label{Eq:Bernoulli_rate_function_lower_bound}
 H(u,p) &\ge&  \frac{(u-p)^2}{2\left(p(1-p)+\frac{1-2p}{3}(u-p) \right)}
   \,\ \eqqcolon \,\ G(u,p) \quad \text{ for }u\in[0,1]\,,
\al
with the denominator of $G(u,p)$ being strictly positive,
and with equality throughout in~\eqref{Eq:Bernoulli_rate_function_lower_bound}
iff $u=p$.
\end{Lem}
\begin{proof} Half the denominator of $G(u,p)$ is a concave quadratic function
$-\frac{1}{3}p^2+\ldots$ of~$p$, hence $>$ the minimum of its limits
at the boundary $\{0,1\}$,
$\min\{\frac{u}{3},\frac{1-u}{3}\}\ge 0$.

We 
put $q\coloneqq 1-p$,
and $h(x)\coloneqq H(p+x,p)= (p+x)\log(1+\frac{x}{p})+(q-x)\log(1-\frac{x}{q})$
and $g(x)\coloneqq G(p+x,p)= \frac{x^2}{2\left(pq +\frac{q-p}{3}x \right)}$
for $x\in [-p,q]$.  For $x\in \mathopen]-p,q\mathclose[$ then
\[
 && h'(x) = \log(1+\tfrac{x}{p}) -\log(1-\tfrac{x}{q})\ , \qquad
    h''(x) = 
     \frac{1}{pq\left(1+\frac{q-p}{pq}x-\frac{x^2}{pq}\right)} \ , \\
 && g'(x) = \frac{\frac{q-p}{6}x^2+pqx}{(pq+\frac{q-p}{3}x )^2} \ , \qquad
    g''(x)= \frac{1}{pq\left(1 +\frac{q-p}{3pq}x\right)^3 } \ ,
\]
and hence $h(0)=h'(0)=0 =g(0)=g'(0)$,
and $h''(x)\ge g''(x)$ due to $h''(x)>0$ and
\[
\frac{1}{pq} \left(\frac{1}{g''(x)} - \frac{1}{h''(x)}\right)
  &=& \frac{(q-p)^2}{3p^2q^2}x^2 + \frac{(q-p)^3}{27p^3q^3}x^3 +\frac{x^2}{pq} \\
  &=& \frac{x^2}{pq}\left( \frac{(q-p)^2}{3pq}\big( 1 + \frac{q-p}{9pq} x  \big)   +1\right)
  \ge 0
\]
with equality throughout iff $x=0$,
using $\min\limits_{x\in[-p,q]} \frac{q-p}{pq}x
= \min\big\{ \frac{q-p}{pq}(-p) , \frac{q-p}{pq}q \big\}
= \min\{\frac{p}{q},\frac{q}{p}\}- 1>-1 > -9$.
Hence $h(x)\ge g(x)$, with equality iff $x=0$.
\end{proof}

\begin{proof}[Proof of Theorem~\ref{Thm:First_bounds_for_Clopper-Pearson_bounds}]
                                            \label{page:proof_of_CP_bounds}
In steps 1 and 2 below,
let $x\in\{1,\ldots,n\}$ and $u\coloneqq \frac{x-1}{n} = \hat{p}-\frac1n$.

\smallskip
1. We use in this step
the notation $F_{n,p},C_{n,p},H(u,p)$ from Theorem~\ref{Thm:Zubkov-Serov}.
For $p\in\mathopen]0,1\mathclose[$ we then have the chain of implications
\la
 p \,<\, \underline{p}_\beta(x)
                      \label{Eq:Implications_for_Clopper-Pearson_upper_bound_upper_bounded}
  &\Leftarrow& \mathrm{B}_{n,p}(\{x,\ldots,n\})  \,<\,  1\!-\!\beta   \\
  &\iff& F_{n,p}(x-1) \,> \, \beta \nonumber \\
  &\Leftarrow& C_{n,p}(x-1) \,\ge\, \beta \nonumber \\
  &\iff& p \,\le\, u \,\text{ and }\, 2H(u,p) \ge \tfrac{z^2}{n} \nonumber \\
  &\Leftarrow&  p \,\le\, u \,\text{ and \eqref{Eq:Quadratic_ineq_CP_1} below } \nonumber
\al
by using in the first step  the definition of $\underline{p}_\beta(x)$
and the isotonicity of $\mathrm{B}_{n,\sdot}(\{x,\ldots,n\})$,
in the third the left hand inequality in~\eqref{Eq:Zubkov_Serov}
in its strict form,
in the fourth the definition of $C_{n,p}$\,, $z\ge 0$ due to $\beta\ge\frac12$\,,
and $H(p,p)=0$\,,
in the fifth Lemma~\ref{Lem:Bernoulli_rate_function_lower_bound}
and the strict positivity of the denominator
in~\eqref{Eq:Bernoulli_rate_function_lower_bound},
where
\la                   \label{Eq:Quadratic_ineq_CP_1}
 (p-u)^2 \,\ge\, \tfrac{z^2}{n}\big( p(1-p) +\tfrac{1-2p}{3}(u-p)  \big)\,.
\al

Inequality~\eqref{Eq:Quadratic_ineq_CP_1} for $p$ is equivalent to
\la                                        \label{Eq:Quadratic_ineq_CP_2}
 Ap^2 +Bp +C &\ge& 0
\al
with
$A\coloneqq 1+\tfrac{z^2}{3n}>0$,
$B\coloneqq -2\, (u+\frac{z^2(1-u)}{3n} )$,
$C\coloneqq u^2 - \tfrac{z^2u}{3n}$,
and is 
implied by taking the left hand root of L.H.S.\eqref{Eq:Quadratic_ineq_CP_2},
\la                                   \label{Eq:Proof_Clopper-Pearson_approximated}
 p&=& \frac{1}{A}\Big( -\frac{B}{2} -\sqrt{ \big(\frac{B}{2}\big)^2 - AC\,}\,\Big)\,,
\al
where
\[
 \big(\frac{B}{2}\big)^2 - AC
  &=& \big(u+\tfrac{z^2(1-u)}{3n} \big)^2 -  \big(1+\tfrac{z^2}{3n}\big)\big( u^2 - \tfrac{z^2u}{3n} \big) \\
  &=& \frac{z^2u(1-u)}{n} + \frac{z^4(1-u(1-u))}{9n^2} \,\ \ge\,\ 0  \,,
\]
and hence
\la    \label{Eq:Proof_Clopper-Pearson_approximated_2}
 \qquad \text{R.H.S.}\eqref{Eq:Proof_Clopper-Pearson_approximated}
  & =& \frac{1}{1+\tfrac{z^2}{3n}}\Big( u+\frac{z^2(1-u)}{3n}
  - \frac{z}{\sqrt{n}} \sqrt{ u(1-u) + \frac{z^2(1-u(1-u))}{9n}  }\  \Big) \,.
\al
Hence if $\text{R.H.S.}\eqref{Eq:Proof_Clopper-Pearson_approximated_2}$
is $\in$ $]0,1[$ and $\le$ $u$, then the
chain~\eqref{Eq:Implications_for_Clopper-Pearson_upper_bound_upper_bounded}
applies to yield
\la                                \label{Eq:Clopper-Pearson_lower_bound_lower_bounded_with_u}
   \text{R.H.S.}\eqref{Eq:Proof_Clopper-Pearson_approximated_2} &<& \underline{p}_\beta(x) \,,
\al
and hence the following observations show
that~\eqref{Eq:Clopper-Pearson_lower_bound_lower_bounded_with_u}
in fact always holds:

In any case,
$\text{R.H.S.}\eqref{Eq:Proof_Clopper-Pearson_approximated_2}
 \le (1+\tfrac{z^2}{3n})^{-1}(u+ \tfrac{z^2}{3n})<1 $ due to $u<1$ and $z\ge0$.
If $\text{R.H.S.}\eqref{Eq:Proof_Clopper-Pearson_approximated_2}\le 0$,
then~\eqref{Eq:Clopper-Pearson_lower_bound_lower_bounded_with_u}
is trivial since $\underline{p}_\beta(x)>0$ due to $x\ge 1$ and $\beta>0$.
So we may assume $\text{R.H.S.}\eqref{Eq:Proof_Clopper-Pearson_approximated_2}  \in\mathopen]0,1\mathclose[$ from now on.
If $z=0$ or $u=0$, then $\text{R.H.S.}\eqref{Eq:Proof_Clopper-Pearson_approximated_2}=u$.
If $z>0$ and $u>0$, then, since also $u<1$,
inequality~\eqref{Eq:Quadratic_ineq_CP_1} is false for $p\coloneqq u$,
and hence $\text{R.H.S.}\eqref{Eq:Proof_Clopper-Pearson_approximated_2}$,
being the left hand root $p$ from \eqref{Eq:Proof_Clopper-Pearson_approximated},
is then $<$ $u$.

Observing that $\text{R.H.S.}\eqref{Eq:Proof_Clopper-Pearson_approximated_2}
    =  \underline{\underline{p}}^{}_\beta(x)$
concludes the proof of the first inequality
in~\eqref{Eq:Clopper-Pearson_lower_bound_lower_bounded}.

\smallskip 2. Inserting $u=\hat{p}-\frac{1}{n}$ and hence
$u(1-u)= (\hat{p}-\frac{1}{n})(\hat{q}+\frac{1}{n}) = \hat{p}\hat{q}+\frac{\hat{p}-\hat{q}}{n}-\frac{1}{n^2}$
yields, with the functions $f_n$ and $g_n$ defined below,
\[
 \underline{\underline{p}}_\beta(x)
  &=& \hat{p} -\tfrac{1}{n} - \frac{1}{1+\tfrac{z^2}{3n}}
  \Big(\tfrac{z^2}{3n}(2\hat{p}-1-\tfrac{2}{n})
  +\frac{z}{\sqrt{n}}\sqrt{\hat{p}\hat{q}  +g_n(\hat{p}) }\,\Big) \\
  &=& \hat{p} -\frac{1}{1+\tfrac{z^2}{3n}}
   \Big(  \tfrac{1}{n}f_n(\hat{p}) + \frac{z}{\sqrt{n}}\sqrt{\hat{p}\hat{q}
     +\tfrac{1}{n}g_n(\hat{p}) }\, \Big)
  \,,
\]
with, for $p\in[0,1]$,
\[
 f_n(p) &\coloneqq& 1+\tfrac{z^2}{3n} +\tfrac{z^2}{3}(2p-1 -\tfrac{2}{n})
 \,\ \le\,\ 1+\tfrac{z^2}{3}\,, \\
 g_n(p) &\coloneqq& 2p-1 -\tfrac{1}{n}+\tfrac{z^2}{9}\big(1-p(1-p)-\tfrac{2p-1}{n}
                     +\tfrac{1}{n^2}\big)\,,
\]
$g_n$ convex and hence
$g(\widehat{p}) \le \max\{g(\frac{1}{n}),g(1)\}$ due to $x\in\{1,\ldots,n\}$,
%
\[
 g_n(1) &=&
    1-\tfrac{1}{n} +\tfrac{z^2}{9}(1 - \tfrac{1}{n}+\tfrac{1}{n^2})
    \,\ \le \,\ 1+\tfrac{z^2}{9}\, , \\
 g_n(\tfrac{1}{n}) &=& \tfrac{1}{n}-1 + \tfrac{z^2}{9} \,\ <\,\ 1+\tfrac{z^2}{9}\, ,
\]
and consequently $ \underline{\underline{p}}_\beta(x)
 \ge \underline{\underline{\underline{p}}}_\beta(x)$.
This concludes the proof of~\eqref{Eq:Clopper-Pearson_lower_bound_lower_bounded}.

\smallskip
3. The inequalities~\eqref{Eq:Clopper-Pearson_upper_bound_upper_bounded}
follow from~\eqref{Eq:Clopper-Pearson_lower_bound_lower_bounded} using
the symmetry~\eqref{Eq:Def_po}.

\smallskip
4. If $x=0$, then $\pu_\beta(x)=0>\puuu_\beta(x)$, since $z\ge 0$.
Analogously, if $x=n$, then $\po_\beta(x)=1<\pooo_\beta(x)$.
Hence~\eqref{Eq:CP_bounds_unrestricted} holds unrestricted as stated.
\end{proof}

{\footnotesize\section*{Notes on the references}

In the present paper, any page numbers in a citation
of a Russian source refer to its English translation. Links 
are provided here only if they lead to the full text for free,
as experienced by us at some point in time. 
The asterisk *  marks 
references
taken from secondary sources, without looking at the original.

}
\end{document}